\documentclass[10pt]{article}
\usepackage{amssymb,amsmath,amsthm,mathrsfs}
\usepackage{epsfig}

\newcommand{\ie}{\emph{i.e.}}

\newcommand{\cf}{\emph{cf}}

\newcommand{\Real}{\mathbb{R}}

\newcommand{\sii}{L^2}

\newcommand{\Dom}{D}

\newcommand{\eps}{\varepsilon}

\newtheorem{Lemma}{Lemma}
\newtheorem{Theorem}{Theorem}
\newtheorem{Corollary}{Corollary}

\theoremstyle{definition}

\newtheorem{ass}{Assumption}

\begin{document}
%
\title{\textbf{\Large
A Hardy inequality in a twisted
Dirichlet-Neumann waveguide
}}
\author{
H.~Kova\v{r}{\'\i}k$^{1,2}$
\ and \
D.~Krej\v{c}i\v{r}\'{\i}k$^{2}$
}
\date{
\footnotesize
\begin{quote}
\begin{itemize}
\item[$^1$]
Institute for Analysis, Dynamics and Modeling, \\
Faculty of Mathematics and Physics, Stuttgart University, \\
PF 80 11 40, D-70569  Stuttgart, Germany
\item[$^2$]
Department of Theoretical Physics,
Nuclear Physics Institute, \\
Academy of Sciences,
250\,68 \v{R}e\v{z} near Prague, Czech Republic
\item[E-mails:]
kovarik@mathematik.uni-stuttgart.de
\ and \
krejcirik@ujf.cas.cz
\end{itemize}
\end{quote}
%
}
\maketitle
\begin{center}
\emph{Dedicated to Pavel Exner on the occasion of his 60th birthday}
\end{center}

\bigskip

\begin{abstract}
\noindent
We consider the Laplacian in a straight strip,
subject to a combination of Dirichlet and Neumann boundary conditions.
We show that a switch of the respective boundary conditions
leads to a Hardy inequality for the Laplacian.
As a byproduct of our method,
we obtain a simple proof of a theorem of
Dittrich and K\v{r}{\'\i}\v{z}~\cite{DKriz1}.
\end{abstract}
%
%
\section{Introduction}
%
The connection between spectral properties of the Laplacian
in a wave\-guide-type domain, the domain geometry
and various boundary conditions
has been intensively studied in the last years,
\cf~\cite{DE,LCM,KKriz} and references therein.
Particular attention has been paid to
the geometrically induced discrete spectrum of the Dirichlet Laplacian
in curved tubes of uniform cross-section \cite{ES,GJ,RB,DE,ChDFK}
or in straight tubes  with a local deformation of the boundary \cite{BGRS,BEGK}.
Roughly speaking, it has been shown that a suitable
bending or a local enlargement of a straight waveguide
represents an effectively attractive perturbation
and leads thus to the presence of eigenvalues
below the essential spectrum of the Laplacian.

On the other hand, recently it has been observed in~\cite{EKK}
that a local rotation of a non-circular cross-section
of a three-dimensional straight tube
creates a kind of repulsive perturbation.
Namely, this type of deformation, called \emph{twist},
gives rise to a Hardy inequality for the Dirichlet Laplacian.
This avoids, up to some extent, the existence of discrete spectrum
in the presence of an additional attractive perturbation,
the bending or local enlargement being two examples.
We refer to~\cite{EKK} for more details
and possible higher-dimensional extensions.

\begin{figure}
\begin{center}
\epsfig{file=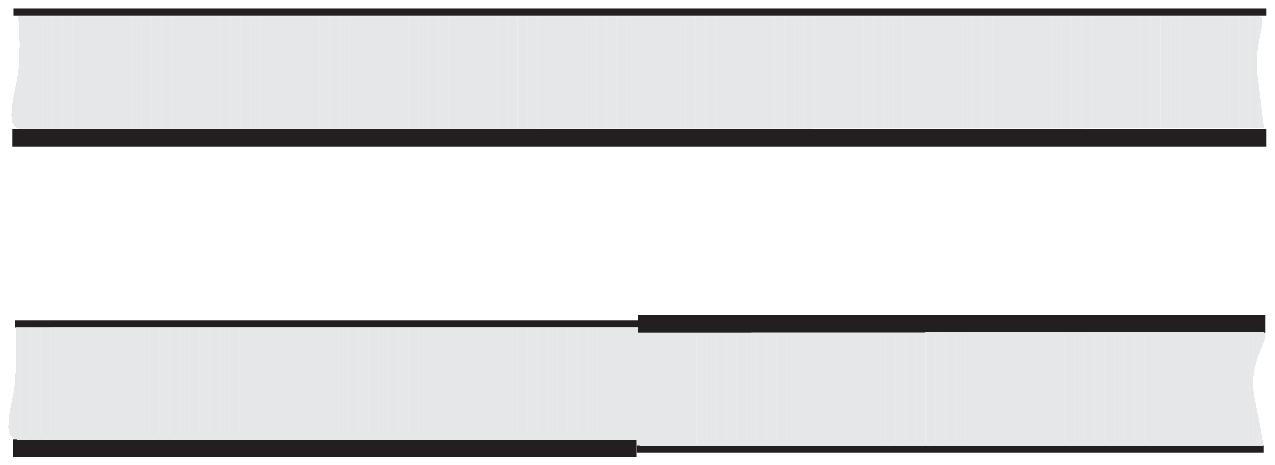,width=0.7\textwidth}
\end{center}
\caption{
We consider the lower waveguide as
a \emph{twist} perturbation of the upper one,
the twist being defined as a switch of Dirichlet (thick lines)
to Neumann (thin lines) boundary conditions at one point,
and \emph{vice versa}.
}\label{Fig.twist}
\end{figure}

\begin{figure}
\begin{center}
\epsfig{file=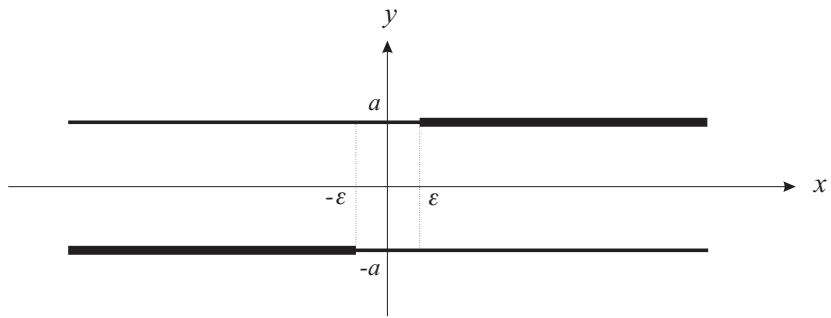,width=0.9\textwidth}
\end{center}
\caption{
The geometry of our waveguide.
The Dirichlet and Neumann boundary conditions
are denoted by thick and thin lines, respectively.
}\label{Fig.strip}
\end{figure}

The purpose of the present note is to demonstrate
an analogous effect of twist in a two-dimensional waveguide
with combined Dirichlet and Neumann boundary conditions.
In this case the twist is represented
by a switch of the boundary conditions
at a given point, \cf~Figure~\ref{Fig.twist}.
More precisely, given a real number~$\eps$ and a positive number~$a$,
let~$-\Delta_\eps$ be the Laplacian
in the strip $\Real \times (-a,a)$,
subject to Dirichlet boundary conditions on
$
  (-\infty,-\eps)\times\{-a\}
  \cup
  (\eps,+\infty)\times\{a\}
$
and Neumann boundary conditions on
$
  (-\eps,+\infty)\times\{-a\}
  \cup
  (-\infty,\eps)\times\{a\}
$,
\cf~Figure~\ref{Fig.strip}.
It can be seen by a simple Neumann bracketing
that the spectrum of~$-\Delta_\eps$
coincides with the interval $[\pi^2/(4a)^2,+\infty)$
for all non-positive~$\eps$.
Our main result shows that for~$\eps$ equal to zero
the operator~$-\Delta_0$ satisfies the
following Hardy type inequality
in the sense of quadratic forms:
\begin{equation} \label{Hardyineq}
  -\Delta_0 - \left(\frac{\pi}{4a}\right)^2
  \ \geq \
  \rho(\cdot)
  \,,
\end{equation}
where $\rho:\Real\times(-a,a) \to \Real$ is a positive function.

We would like to emphasize that in the situation where the boundary
conditions are not exchanged
-- \ie~the Laplacian in $\Real\times(-a,a)$
with uniform Dirichlet boundary conditions on one connected
part of the boundary and Neumann boundary conditions on the other one,
\cf~the upper waveguide in Figure~\ref{Fig.twist} --
the essential spectrum coincides with the essential spectrum of our
waveguide, but the inequality~(\ref{Hardyineq}) fails to hold
for any non-trivial $\rho \geq 0$.
The latter can be shown by a simple test-function argument.
In other words, the switch of the boundary conditions
creates a kind of repulsive perturbation
represented by the function~$\rho$.
This leads to a certain stability of the spectrum
similar to the one observed in~\cite{EKK}.
In particular, it follows from~(\ref{Hardyineq})
that the discrete spectrum remains empty
after perturbing~$-\Delta_0$
by a sufficiently small attractive perturbation.

One example of attractive perturbation is changing the boundary
conditions by increasing the parameter~$\eps$,
\cf~Figure~\ref{Fig.strip}.
Due to the switch of the boundary conditions,
the discrete eigenvalues do not appear for any positive~$\eps$,
but only when~$\eps$ exceeds certain critical value $\eps_c>0$.
This effect was already observed
by Dittrich and K\v{r}{\'\i}\v{z} in~\cite{DKriz1}.
Their result is obtained by a tedious decomposition of the
Laplacian into the ``transverse basis'' and this also provides an estimate
on the critical value~$\eps_c$ for which the eigenvalues emerge
from the essential spectrum:
\begin{equation}\label{estimates}
  0.16\,a
  < \eps_c <
  0.68\,a
  \,.
\end{equation}
Since the proof of our Hardy inequality~(\ref{Hardyineq})
can be easily carried over to the case when~$\eps$ is positive and small
enough, we get as a byproduct of our method
an alternative estimate on $\eps_c$, too.
The latter is worse than the one presented in~\cite{DKriz1},
but on the other hand much simpler to obtain.

Finally, let us mention that Hardy inequalities for Schr\"odinger
operators in two dimensions can be achieved by adding an appropriate
local magnetic field to the system, too.
This was first observed in~\cite{LW} and later modified in~\cite{EK}
for Schr\"odinger operators in waveguides, \cf~also~\cite{BEK}.
Curved waveguides in a homogeneous magnetic field have been
recently studied in~\cite{OM}.

\section{Main results and ideas}\label{Sec.Main}
%
The Laplacian~$-\Delta_\eps$ is defined as the unique
self-adjoint operator associated with the closure
of the quadratic form~$Q_\eps$ defined
in $\sii\big(\Real\times(-a,a)\big)$ by
\begin{equation}\label{form}
  Q_\eps[\psi]
  := \int_{\Real \times (-a,a)}
  \left(
  |\partial_1\psi(x,y)|^2 +
  \partial_2\psi(x,y)|^2
  \right)
  dx\,dy
\end{equation}
and by the domain $\Dom(Q_\eps)$ which consists of restrictions
to $\Real\times(-a,a)$ of infinitely smooth functions
with compact support in~$\Real^2$
and vanishing on the part of the boundary
where the Dirichlet boundary conditions are imposed
(\cf~\cite{DKriz1} for more details).
We are interested in the shifted quadratic form~$\tilde{Q}_\eps$
defined on the form domain~$\Dom(Q_\eps)$ by the prescription
\begin{equation}\label{form.shifted}
  \tilde{Q}_\eps[\psi]
  := Q[\psi]
  - \left(\frac{\pi}{4a}\right)^2
  \int_{\Real \times (-a,a)} |\psi(x,y)|^2 \, dx\,dy
  \,.
\end{equation}

If~$\eps$ is negative,
so that the opposite Dirichlet boundary conditions overlap,
one can estimate the second term in~(\ref{form})
by the lowest eigenvalue of the Laplacian in the cross-section
of length~$2a$,
subject to Dirichlet-Dirichlet
or Dirichlet-Neumann boundary conditions.
Neglecting the first term in~(\ref{form}),
this immediately yields
\begin{equation}\label{negative}
  -\Delta_\eps - \left(\frac{\pi}{4a}\right)^2
  \ \geq \
  3 \left(\frac{\pi}{4a}\right)^2 \
  \chi_{(\eps,-\eps) \times (-a,a)}(\cdot)
  \qquad\mbox{if}\qquad
  \eps < 0
\end{equation}
in the sense of quadratic forms.
Here~$\chi_M$ denotes the characteristic function of a set~$M$.
The right hand side provides a non-negative Hardy weight in this case.

Of course, the trivial estimate leading to~(\ref{negative})
is not useful for non-negative~$\eps$, in which case other
methods have to be used. In this paper we get:

\begin{Theorem}\label{Thm}
Given a real number~$\eps$ and a positive number~$a$,
let $-\Delta_\eps$ be the Laplacian
in the strip $\Real\times(-a,a)$,
subject to Dirichlet boundary conditions on
$
  (-\infty,-\eps)\times\{-a\}
  \cup
  (\eps,+\infty)\times\{a\}
$
and Neumann boundary conditions on
$
  (-\eps,+\infty)\times\{-a\}
  \cup
  (-\infty,\eps)\times\{a\}
$.
\begin{itemize}
\item[(i)]
There exists a positive constant~$c$
such that the inequality
\begin{equation} \label{localhardy}
  -\Delta_0 - \left(\frac{\pi}{4a}\right)^2
  \ \geq \
  c \, \chi_\omega(\cdot)
\end{equation}
holds in the sense of quadratic forms.
Here $\omega \supseteq (-a,a)\times(-a,a)$
and
$$
  c \geq s_1 \left(\frac{\pi}{4a}\right)^2
  ,
$$
where~$s_1$ is the smallest root of the equation
\begin{equation}\label{root.Hardy}
  \sqrt{1-s} \ \tanh\left(
  \frac{\pi\,\sqrt{1-s}}{2\sqrt{2}}
  \right)
  = \sqrt{1/2+s} \ \tan\left(
  \frac{\pi\,\sqrt{1/2+s}}{2\sqrt{2}}
  \right) .
\end{equation}
\item[(ii)]
There exists a positive constant $\eps_c \geq t_1 \;\! a$
such that
$$
  \sigma(-\Delta_\eps) = \big[\pi^2/(4a)^2,\infty\big)
$$
for all $\eps \leq \eps_c$.
Here~$t_1$ is the smallest positive root of the equation
\begin{equation}\label{root.eps}
  \tanh\left(
  \frac{\pi\,(1-t)}{2\sqrt{2}}
  \right)
  = \sqrt{1/2} \ \tan\left(
  \frac{\pi\,(1+t)}{2\sqrt{2}}
  \right) .
\end{equation}
\end{itemize}
\end{Theorem}

The first result, \ie~the Hardy inequality for~$-\Delta_0$, is new.
On the other hand, a positive lower bound on~$\eps_c$
has already been established in~\cite{DKriz1}, \cf~(\ref{estimates}).
In~\cite{DKriz1} the authors also find the numerical value
$
  \eps_c \approx 0.52\,a
$.
We have $s_1 \approx 0.039$ and $t_1 \approx 0.061$,
and these numbers cannot be much improved by our method
(\cf~the end of Section~\ref{Sec.1D} for more details).

\begin{figure}
\begin{center}
\epsfig{file=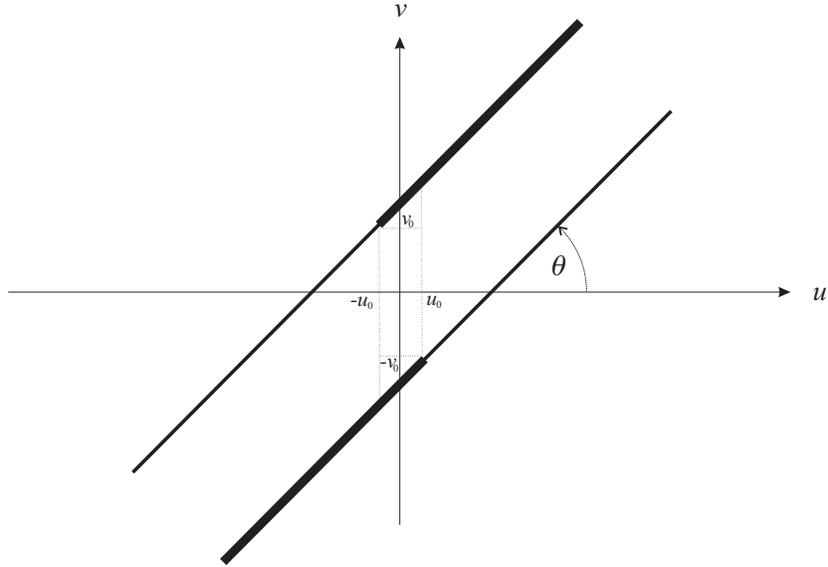,width=0.9\textwidth}
\end{center}
\caption{
Rotating the Cartesian coordinate system by an appropriate angle~$\theta$,
one can employ the repulsive interaction
due to the proximity of opposite Dirichlet boundary conditions
(thick lines).
}\label{Fig.rotated}
\end{figure}

Although the effect which causes~(\ref{localhardy})
is very similar to the twist studied in~\cite{EKK},
the methods used in the respective proofs are completely different.
The reason is that in our case the twist represents
a singular deformation in the sense that
it is discontinuous and occurs at one point only.
Our main idea to prove Theorem~\ref{Thm}
is to introduce rotated Cartesian coordinates
in which one can employ the repulsive interaction
due to the proximity of opposite Dirichlet boundary conditions,
\cf~Figure~\ref{Fig.rotated}.
This is done in Section~\ref{Sec.reduction}
where the initial problem is reduced
to an ordinary differential equation.
The latter is then investigated in Section~\ref{Sec.1D}
by standard methods for one-dimensional Schr\"odinger
operators.

Note that Theorem~\ref{Thm} contains a weaker version of
inequality~(\ref{Hardyineq}), namely with a compactly supported Hardy
weight. However, (\ref{Hardyineq})~can be easily deduced from it:
\begin{Corollary}\label{Corol}
Inequality~(\ref{Hardyineq}) holds true
with the function~$\rho$ given by
\begin{equation*}
  \rho(x,y) := \frac{c_h}{1+x^2}
  \,, \qquad
  c_h := \left(
  \max\big\{
  16,
  c^{-1}(2+16/a^2)
  \big\}
  \right)^{-1}
  \,,
\end{equation*}
where $c$ is the constant from Theorem~\ref{Thm}.
\end{Corollary}
A short proof of Corollary~\ref{Corol},
based on the classical one-dimensional Hardy inequality,
is given in the concluding Section~\ref{Sec.Corol}.

\section{Reduction to a one-dimensional problem}\label{Sec.reduction}
%
Hereafter we consider non-negative~$\eps$ only.
Let $(x,y) \in \Real\times(-a,a)$.
We introduce rotated Cartesian coordinates $(u,v)$
by the change of variables
\begin{equation}\label{change}
  (x,y) = f(u,v)
  := \big(
  u \cos\theta + v \sin\theta,
  - u \sin\theta + v \cos\theta
  \big)
  \,,
\end{equation}
where $\theta\in(0,\pi/2)$.
Clearly, the mapping $f:\Omega\to\Real\times(-a,a)$
is a diffeomorphism with the preimage
\begin{align*}
  \Omega:=f^{-1}(\Real \times I)
  &= \left\{
  (u,v) \in \Real^2 \, | \ u_-(v) < u < u_+(v)
  \right\}
  \\
  &= \left\{
  (u,v) \in \Real^2 \, | \ v_-(u) < v < v_+(u)
  \right\}
  \,,
\end{align*}
where
$$
  u_\pm(v) := \frac{\pm a + v\cos\theta}{\sin\theta}
  \,,
  \qquad
  v_\pm(u) := \frac{\pm a + u\sin\theta}{\cos\theta}
  \,.
$$

Introducing the (unitary) change of trial function
$
  \psi \mapsto \psi \circ f := \phi
$
into the functional~(\ref{form}), we find
\begin{equation}\label{form.rotated}
  Q_\eps[\phi \circ f^{-1}]
  = \int_\Omega
  \left(
   |\partial_1\phi(u,v)|^2 +
  \partial_2\phi(u,v)|^2
  \right) du \, dv
  \,.
\end{equation}
From the formulae
$$
  \phi\big(u,v_\pm(u)\big)
  = \psi\left(
  \frac{u \pm a\sin\theta}{\cos\theta},
  \pm a
  \right) ,
  \qquad
  \phi\big(u_\pm(v),v\big)
  = \psi\left(
  \frac{v \pm a\cos\theta}{\sin\theta},
  \mp a
  \right) ,
$$
we observe the two following properties, respectively.
First, $v\mapsto\phi(u,v)$ with~$u$ fixed
satisfies Dirichlet boundary conditions
at both boundary points $v_\pm(u)$
if, and only if,
\begin{equation}\label{u0}
  |u| < u_0 := a\sin\theta - \eps\cos\theta
  \,;
\end{equation}
otherwise it satisfies a combination
of Dirichlet and (generalized) Neumann boundary conditions.
Second, $u\mapsto\phi(u,v)$ with~$v$ fixed
satisfies a combination of Dirichlet
and (generalized) Neumann boundary conditions,
if, and only if,
\begin{equation}\label{v0}
  |v| > v_0 := a\cos\theta + \eps\sin\theta
  \,;
\end{equation}
otherwise it satisfies (generalized)
Neumann boundary conditions (\ie~none).
While~$v_0$ is positive by definition,
we need to assume that
\begin{equation}\label{eps.bound}
  \eps < a \, \tan\theta
\end{equation}
in order to ensure the positivity of~$u_0$.

We proceed by estimating
the form~(\ref{form.rotated}) as follows.
We estimate the second term in~(\ref{form.rotated})
by the lowest eigenvalue of the Laplacian
in the cross-section of length
$
  v_+(u) - v_-(u) = 2a / \cos\theta
$,
subject to the boundary conditions
of the type that $v\mapsto\phi(u,v)$ satisfies.
We also estimate the first term in~(\ref{form.rotated})
by the lowest eigenvalue of the Laplacian
in the cross-section of length
$
  u_+(v) - u_-(v) = 2a / \sin\theta
$,
subject to the boundary conditions
of the type that $u\mapsto\phi(u,v)$ satisfies,
but only in the subset of~$\Omega$
where $|u|>u_0$ and $|v|>v_0$.
That is,
\begin{equation}\label{principal.bound}
  \tilde{Q}_\eps[\phi \circ f^{-1}]
  \geq
  \int_{\Omega_1 \cup \Omega_2}
  |\partial_1\phi|^2
  + q_+ \int_{\Omega_1} |\phi|^2
  - q_- \int_{\Omega_2} |\phi|^2
  \,,
\end{equation}
where
\begin{equation*}
  \Omega_1
  := \left\{
  (u,v) \in \Omega \, | \
  |u| < u_0
  \right\}
  \,,
  \qquad
  \Omega_2
  := \left\{
  (u,v) \in \Omega \, | \
  |v| < v_0 , \
  |u| > u_0
  \right\}
  \,,
\end{equation*}
and
\begin{equation}\label{q}
  q_+ := \left(\frac{\pi}{4a}\right)^2 (4\cos^2\theta-1)
  \,,
  \qquad
  q_- := \left(\frac{\pi}{4a}\right)^2 \sin^2\theta
  \,.
\end{equation}
Hereafter we further restrict the angle~$\theta$
by the requirement
\begin{equation}\label{condition}
  \theta \in (0,\pi/3)
  \,,
\end{equation}
so that the term~$q_+$ is positive.

We use the intermediate bound~(\ref{principal.bound})
as the starting point of the reduction to a one-dimensional problem.
Let us introduce the disjoint sets
\begin{equation*}
  \Omega_1'
  := \left\{
  (u,v) \in \Omega \, | \
  |u| < u_0 , \
  |v| > v_0
  \right\}
  \,,
  \qquad
  \Omega_2'
  := \left\{
  (u,v) \in \Omega \, | \
  |v| < v_0
  \right\}
  \,,
\end{equation*}
and note that the inclusions
$
  \Omega_1' \subset \Omega_1
$
and
$
  \Omega_2' \subset \Omega_1\cup\Omega_2
$
hold.
Consequently, under the assumption~(\ref{condition}),
(\ref{principal.bound})~implies the cruder bound
\begin{equation}\label{crude}
  \tilde{Q}_\eps[\phi \circ f^{-1}]
  \geq
  q_+ \int_{\Omega_1'} |\phi(u,v)|^2 \, du \, dv
  + \int_{\Omega_2'}
  \lambda(v) \, |\phi(u,v)|^2
  \, du \, dv
  \,,
\end{equation}
where $\lambda(v) \in (q_-,q_+)$ is the lowest eigenvalue
of the one-dimensional Neumann Schr\"odinger operator
with the step-like potential
$$
  V(u,v) :=
  q_+ \ \chi_{(-u_0,u_0)}(u)
  - q_- \ \chi_{(u_-(v),-u_0)\cup(u_0,u_+(v))}(u)
  \,.
$$
More precisely,
\begin{equation}\label{lambda}
  \lambda(v) :=
  \inf_{\varphi}
  \frac{\int_{u_-(v)}^{u_+(v)}
  \Big[
  |\varphi'(u)|^2
  + V(u,v) \, |\varphi(u)|^2
  \Big] \, du}
  {\int_{u_-(v)}^{u_+(v)} |\varphi(u)|^2 \, du}
  \,,
\end{equation}
where the infimum is taken over all non-zero
functions from the Sobolev space
$W^{1,2}\big(u_-(v),u_+(v)\big)$.

The formula~(\ref{crude}) together with~(\ref{lambda})
transfers the initial two-dimensional problem
into the study of an ordinary differential equation.
That is, it remains to investigate
the function $v\mapsto\lambda(v)$.

\section{Study of the one-dimensional problem}\label{Sec.1D}
\begin{figure}
\begin{center}
\epsfig{file=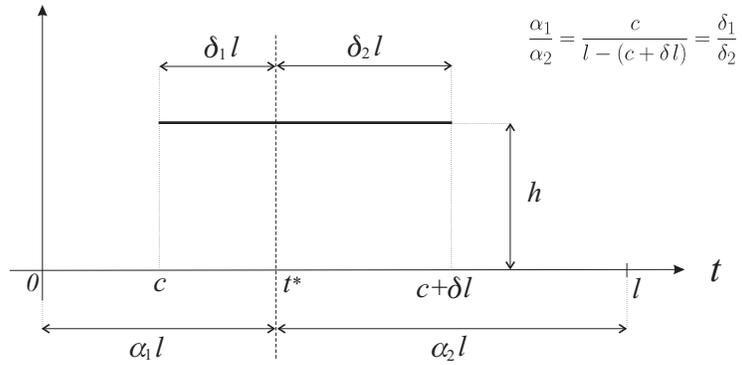,width=0.8\textwidth}
\end{center}
\caption{
The construction used in the proof of Lemma~\ref{Lemma}.
}\label{Fig.1D}
\end{figure}

%
\label{proof}
First of all, we observe that $v\mapsto\lambda(v)$
is an even function with values in the open interval $(q_-,q_+)$
due to~(\ref{condition}).
Furthermore, its minimum is attained
at the boundary points $v=\pm v_0$:
\begin{Lemma}\label{Lemma}
One has
$
{\displaystyle
  \inf_{v\in(-v_0,v_0)} \lambda(v)
  = \lambda(v_0)
  \,.
}
$
\end{Lemma}
\begin{proof}
Let~$h$, $l$ and~$\delta$ be positive numbers such that $\delta<l$.
For any real~$c$,
we consider the one-dimensional Schr\"odinger operator
\begin{equation*}
  H_c := -\Delta + h \, \chi_{(c,c+\delta l)}
  \qquad\qquad\mbox{in}\qquad
  \sii\big((0,l)\big)
  \,,
\end{equation*}
subject to Neumann boundary conditions.
($H_c$ is introduced in a standard way
through the associated quadratic form
defined in $W^{1,2}((0,l))$.)
Let us show that
\begin{equation}\label{to.show}
  \forall c\in(0,l-\delta l), \qquad
  \inf\sigma(H_c) \geq \inf\sigma(H_0)
  \,,
\end{equation}
which is equivalent to the statement of the Lemma.

The reader is advised to consult Figure~\ref{Fig.1D}
for the following construction.
Given $c\in(0,l-\delta l)$,
we find $\alpha_1,\alpha_2 \in (0,1)$
such that
\begin{equation*}
  \alpha_1+\alpha_2=1
  \qquad\mbox{and}\qquad
  \frac{\alpha_1}{\alpha_2}= \frac{c}{l-(c+\delta l)}
  \,.
\end{equation*}
We also define parameters $\delta_1,\delta_2 \in (0,\delta)$
by the equations
\begin{equation*}
  \delta_1+\delta_2 = \delta
  \qquad\mbox{and}\qquad
  \frac{\delta_1}{\delta_2} = \frac{\alpha_1}{\alpha_2}
  \,.
\end{equation*}
It follows that
$
  \alpha_1 \;\! l
  = c + \delta_1 l
$.
Let $t^*:=\alpha_1 \;\! l \in (0,l)$.

The minimax principle yields
\begin{equation*}
  \inf\sigma(H_c) \geq \inf\sigma(H_c^N)
  \,,
\end{equation*}
where~$H_c^N$ is the operator obtained from~$H_c$
by imposing an additional Neumann boundary condition
at the point~$t^*$.
$H_c^N$~is a direct sum of two operators,
which are unitarily equivalent to
\begin{align*}
  T_1 &:= -\Delta + h \, \chi_{(0,\delta_1 l)}
  &\mbox{in}&\qquad
  \sii\big((0,\alpha_1\;\!l)\big) \,,
  \\
  T_2 &:= -\Delta + h \, \chi_{(0,\delta_2 l)}
  &\mbox{in}&\qquad
  \sii\big((0,\alpha_2\;\!l)\big) \,,
\end{align*}
respectively,
both subject to Neumann boundary conditions.
Hence,
\begin{equation}\label{spectrum}
  \sigma(H^N_c) = \sigma(T_1)\cup\sigma(T_2)
  \,.
\end{equation}
Obvious changes of variable show that
that~$T_1$ and~$T_2$ are unitarily equivalent
to the operators
\begin{align*}
  \hat{T}_1 &:= - (\delta/\delta_1)^2 \, \Delta
  + h \, \chi_{(0,\delta l)}
  &\mbox{in}&\qquad
  \sii\big((0,l)\big) \,,
  \\
  \hat{T}_2 &:= - (\delta/\delta_2)^2 \, \Delta
  + h \, \chi_{(0,\delta l)}
  &\mbox{in}&\qquad
  \sii\big((0,l)\big) \,,
\end{align*}
respectively,
both subject to Neumann boundary conditions.
Consequently,
$$
  \hat{T}_1 \geq H_0
  \qquad\mbox{and}\qquad
  \hat{T}_2 \geq H_0
$$
in the sense of quadratic forms.
This together with~(\ref{spectrum}) implies~(\ref{to.show}).
\end{proof}

As a consequence of~(\ref{crude}) and the above Lemma,
we therefore obtain
\begin{equation}\label{almost}
  \tilde{Q}_\eps[\phi \circ f^{-1}]
  \geq
  \lambda(v_0)
  \int_{\Omega_1' \cup \Omega_2'} |\phi(u,v)|^2
  \, du \, dv
  \,.
\end{equation}

We now turn to a more quantitative study of $\lambda(v_0)$.
The eigenvalue problem associated with~(\ref{lambda})
can be solved explicitly in the intervals
where the potential~$V$ is constant.
Matching these solutions in the discontinuity points of~$V$,
one easily finds that $\lambda(v_0)$ coincides with
the smallest root~$\lambda\in(q_-,q_+)$ of the equation
\begin{equation}\label{implicit}
  g_1(\lambda,\eps,\theta)
  = g_2(\lambda,\eps,\theta)
  \,,
\end{equation}
where
\begin{align*}
  g_1(\lambda,\eps,\theta)
  &:= \sqrt{q_+-\lambda} \ \tanh\left(
  2 u_0 \sqrt{q_+-\lambda}
  \right) ,
  \\
  g_2(\lambda,\eps,\theta)
  &:= \sqrt{q_-+\lambda} \ \tan\left(
  2 v_0 \cot\theta \, \sqrt{q_-+\lambda}
  \right) .
\end{align*}
Recall that $q_+,q_-$ and~$u_0,v_0$
are introduced in~(\ref{q}) and~(\ref{u0})--(\ref{v0}), respectively.
Of course, $g_2$~is not defined for all the values
of the parameters~$\lambda,\eps,\theta$,
and we should rather multiply~(\ref{implicit}) by
$
  \cos\big( 2 v_0 \cot\theta \sqrt{q_-+\lambda} \big)
$,
but the resulting (regular) equation cannot
be satisfied if the cosine equals zero,
so we can leave~(\ref{implicit}) in the present form.

Let us first consider the case $\eps=0$.
A necessary condition to guarantee
the eligibility of our method to prove Theorem~\ref{Thm}
is that $\lambda(v_0)$ is positive
for certain angle~$\theta$ satisfying~(\ref{condition}).
A numerical study of~(\ref{implicit}) shows
that $\lambda(v_0)$ achieves its maximum,
given approximately by
$
  0.040 \, \pi^2/(4a)^2
$,
for the angle
$
  \theta \approx 0.774
$.
Observing that the optimal angle is close to
$
  \pi/4 \approx 0.785
$,
let us fix henceforth:
\begin{equation}\label{choice}
  \theta = \pi/4
  \,.
\end{equation}
Since $\lambda \mapsto g_1(\lambda,0,\pi/4)$
is decreasing and continuous,
$\lambda \mapsto g_2(\lambda,0,\pi/4)$
is increasing and continuous,
and at $\lambda = 0$ we have
\begin{equation}\label{fraction}
  \frac{g_1(0,0,\pi/4)}{g_2(0,0,\pi/4)}
  = \sqrt{2} \, \tanh\big(\sqrt{2}\,\pi/4\big)
  > 1
  \,,
\end{equation}
it follows that $\lambda(v_0)$ is indeed positive
for the choice~(\ref{choice}).
As for the numerical value,
it is straightforward to check that~(\ref{implicit})
reduces to~(\ref{root.Hardy})
and we find that the smallest root~$s_1$ of the latter
equals approximately $0.039$.
Summing up, (\ref{almost})~implies
\begin{equation*}
  \tilde{Q}_\eps[\phi \circ f^{-1}]
  \geq
  s_1 \left(\frac{\pi}{4a}\right)^2
  \int_{\Omega_1'\cup\Omega_2'} |\phi(u,v)|^2
  \, du \, dv
  \,,
\end{equation*}
provided the angle~$\theta$
is chosen according to~(\ref{choice}).
In order to establish~(i) of Theorem~\ref{Thm},
it remains to realize that
$$
  \overline{f\big(\Omega_1'\cup\Omega_2'\big)}
  \supset (-a,a)\times(-a,a)
  \,,
$$
where~$f$ is given by~(\ref{change}).

In the case of positive~$\eps$,
we put~$\lambda$ equal to zero in~(\ref{implicit})
and look for the smallest positive~$\eps$
satisfying the equation~(\ref{implicit}).
This root satisfies the restriction~(\ref{eps.bound})
because $\eps \mapsto g_1(0,\eps,\pi/4)$
is decreasing and continuous,
$\eps \mapsto g_2(0,\eps,\pi/4)$
is increasing and continuous,
$g_1(0,a,\pi/4)=0$,
$g_2(0,\eps,\pi/4)$ tends to $+\infty$ as $\eps \to a$,
and we have~(\ref{fraction}) for $\eps=0$.
It is straightforward to check that~(\ref{implicit})
reduces to~(\ref{root.eps}) for the choice~(\ref{choice})
and the smallest positive root~$t_1$ of the latter
equals approximately $0.061$.
Again, a more detailed numerical study of~(\ref{implicit})
shows that the best result reachable by the present method
gives $\eps_c \approx 0.063 \, a$
with the optimal angle
$
  \theta \approx 0.759
$.

This concludes the proof of Theorem~\ref{Thm}.

\section{Proof of Corollary~\ref{Corol}}\label{Sec.Corol}
%
The local Hardy inequality~(\ref{localhardy})
is equivalent to
\begin{equation*}
  \int_{(-a,a)\times(-a,a)} |\psi|^2
  \leq \int_{\Real\times(-a,a)} |\partial_1\psi|^2
  + \int_{\Real\times(-a,a)} |\partial_2\psi|^2
  - \left(\frac{\pi}{4a}\right)^2
  \int_{\Real\times(-a,a)} |\psi|^2
\end{equation*}
for any
$
  \psi\in\overline{\Dom(Q_\eps)}
  \subset
  W^{1,2}\big(\Real\times(-a,a)\big)
$.
Here the sum of the last two terms on the right hand side
is non-negative due to the boundary conditions
that~$\psi$ satisfies.
Consequently, Corollary~\ref{Corol} follows at once
by means of the following Hardy-type inequality
for a Schr\"odinger operator in a strip
with the potential being a characteristic function:
\begin{Lemma}\label{Lem.Hardy}
For any $\psi\in W^{1,2}\big(\Real\times(-a,a)\big)$,
\begin{equation*}
  \int_{\Real\times(-a,a)}
  w^{-2} \, |\psi|^2
  \ \leq \
  16 \int_{\Real\times(-a,a)} |\partial_1\psi|^2
  + \big(2+64/|J|^2\big) \int_{J\times(-a,a)} |\psi|^2
  \,,
\end{equation*}
where
$
  w(x,y) := \sqrt{1+(x-x_0)^2}
$,
$J$~is any bounded subinterval of~$\Real$
and~$x_0$ is the mid-point of~$J$.
\end{Lemma}
\noindent
This Lemma can be established quite easily
by means of the classical one-dimensional Hardy inequality
$
  \int_{\Real} x^{-2} |v(x)|^2 \, dx
  \leq 4 \int_{\Real}\,|v'(x)|^2\, dx
$
valid for any $v\in W^{1,2}(\Real)$ with $v(0)=0$
and Fubini's theorem;
we refer the reader to \cite[Sec.~3.3]{EKK}
or \cite[proof of Lem.~2]{K3}
for more details.

\section*{Acknowledgement}
%
The work has partially been supported by
the Czech Academy of Sciences and its Grant Agency
within the projects IRP AV0Z10480505 and A100480501,
and by DAAD within the project D-CZ~5/05-06.

%
%
\providecommand{\bysame}{\leavevmode\hbox to3em{\hrulefill}\thinspace}
\providecommand{\MR}{\relax\ifhmode\unskip\space\fi MR }
\providecommand{\MRhref}[2]{%
  \href{http://www.ams.org/mathscinet-getitem?mr=#1}{#2}
}
\providecommand{\href}[2]{#2}

%

\newpage
%
\end{document}